# Diseño en FPGA de un turbo decodificador para cdma2000

*FPGA design of a cdma2000 turbo decoder*

Maribell Sacanamboy Franco[1], Fabio G. Guerrero[2]

**RESUMEN**

En este trabajo se presenta el diseño hardware en FPGA de un turbo decodificador para el estándar cdma2000. El trabajo incluye un estudio y análisis matemático del proceso de turbo decodificación, basado en el algoritmo MAX-Log-MAP. Se presentan los resultados de decodificación obtenidos para un tamaño de paquete de doscientos cincuenta bits, un análisis de área y desempeño y se identifican las variables fundamentales para el diseño hardware en la turbo decodificación.

Palabras clave: turbo decodificación, cdma2000, turbo códigos, MAX-Log-MAP, diseño en FPGA.

*ABSTRACT*

*This paper presents the FPGA hardware design of a turbo decoder for the cdma2000 standard. The work includes a study and mathematical analysis of the turbo decoding process, based on the MAX-Log-MAP algorithm. Results of decoding for a packet size of two hundred fifty bits are presented, as well as an analysis of area versus performance, and the key variables for hardware design in turbo decoding.*

## INTRODUCCIÓN

La invención de la turbo codificación hacia finales del siglo XX probó la existencia práctica de un esquema de codificación que permitía obtener una probabilidad de error arbitrariamente pequeña para cualquier velocidad de transmisión de información menor o igual a la capacidad de un canal con ruido blanco gausiano aditivo, tal como lo había demostrado matemáticamente Shannon en 1948. Entre los beneficios prácticos de un sistema de codificación óptimo se encuentra el poder transmitir información lo más rápido posible sin utilizar más potencia de la estrictamente necesaria para una BER (Bit Error Rate) dada. Por esta razón los turbo códigos han encontrado aplicación práctica en diferentes estándares de comunicaciones actuales tales como UMTS, CDMA2000, DVB-RCS, DVB-RCT, Inmarsat, Eutelsat, WiMAX, CCSDS (deep space) y otros [1].

Los turbo códigos, en su forma original, se encuentran compuestos de dos códigos convolucionales sistemáticos recursivos separados por un entrelazador, el cual tiene la función de mezclar los datos de entrada hacia el segundo codificador con el fin de aumentar la distancia entre los vectores enviados al receptor. Para lograr el límite de la capacidad del canal la turbo decodificación debe emplear decisiones probabilísticas de tipo suave, es decir, hacer uso de la información de la confiabilidad sobre el valor de la decisión realizada por el modulador acerca de los símbolos recibidos. La palabra turbo se refiere al intercambio iterativo de información entre los decodificadores constituyentes para mejorar el nivel de corrección de errores al aumentar progresivamente el número de iteraciones [2].

Mientras, desde el punto de vista hardware, el diseño del turbo codificador en el transmisor es muy sencillo, el diseño del turbo decodificador en el receptor es bastante complejo.

En [3] se reporta el diseño de una arquitectura reconfigurable para diferentes estándares basada en un multiprocesador el cual proporciona un alto rendimiento y un consumo de potencia escalable, sin embargo las especificaciones y detalles del diseño no están dados.

En [4] se reporta el diseño de un procesador SIMD para un turbo decodificador que soporta múltiples estándares implementado en VLSI, sin embargo los detalles del diseño no son muy descriptivos.

En [5] se reporta el diseño de un turbo decodificador en una FPGA el cual procesa datos de entrada a una tasa de 600Mb/s, utilizando un diseño novedoso pero no se describe en detalle la arquitectura propuesta.

[1] Departamento de Ciencias de la ingeniería y la computación. Pontificia Universidad Javeriana Cali, calle 18 No. 118-250 vía Pance, Cali, Colombia (Sur América); msacanamboy@javerianacali.edu.co
[2] Escuela de ingeniería eléctrica y electrónica. Universidad del Valle Cali, Ciudad universitaria Meléndez, calle 13 100-00, Edificio 355, oficina 2005, Cali Colombia (Sur América). Fabio.guerrero@correounivalle.edu.co

En [6] se presenta una implementación de un turbo decodificador para cdma2000, basado en una modificación del algoritmo SOVA e implementado sobre una FPGA, a pesar de que en las conclusiones se describe que este desarrollo alcanza un desempeño más alto y una latencia más baja, que un decodificador comercial, en su arquitectura no se muestran los detalles de su diseño, está más orientado a diagramas de bloques.

En [7] se reporta el diseño hardware de un turbo decodificador basado en bloques, lo cual hace de muy poca utilidad a la hora de plantear un diseño original, dado que no hay una descripción detallada del diseño.

En [8] se presenta un turbo decodifcador de alta velocidad basado en Max-Log-MAP, en cual se muestra algunos bloques en detalle de la arquitectura, sin embargo no se muestra la arquitectura completa del turbo decodificador especificando por ejemplo tamaño de buses, señales de control y ampliando en detalle algunos otros bloques como el de normalización el cual usan para evitar desbordamiento en el cálculo de las métricas.

El objetivo de este artículo es presentar el diseño hardware en detalle de un sistema de turbo codificación y decodificación en un sistema práctico, tomando como ejemplo la especificación del estándar cdma2000 [9].

El algoritmo de decodificación usado en este trabajo está basado en Max-Log-MAP, el cual es una variante del algoritmo MAP (Máximo A Posteriori) [10].

Entender las variables que determinan el desempeño del diseño hardware de un sistema de turbo codificación es algo muy importante debido a las limitaciones intrínsecas del proceso de decodificación [11].

En este artículo se muestra en detalle cada uno de los componentes del sistema usando hardware FPGA (Field Programmable Gate Arrays), se explica cómo interaccionan los componentes entre sí, y se identifican los principales aspectos del diseño hardware de la turbo decodificación. Este trabajo permite conocer en profundidad las propiedades propias del diseño hardware de turbo codificación.

Este trabajo está organizado de la siguiente manera: en la siguiente sección se presenta la metodología utilizada para esta investigación, luego se presentan los resultados más relevantes obtenidos, una discusión sobre los mismo y, finalmente, se sintetizan los principales aportes y conclusiones de este trabajo.

## METODOLOGIA

Los diseños presentados en este trabajo fueron realizados usando descripción estructural en VHDL y captura esquemática. Las fases de síntesis y simulación se realizaron Quartus II versión 9 y el diseño fue sintetizado en la FPGA Stratix III EP3SE80F1152C2.

Dado que durante todo el proceso de contaminación y decodificación se emplean datos suaves, los datos se normalizaron por 1024, quedando entonces con una representación binaria de 20 bits y los datos negativos con complemento a dos. Basado en esta normalización y trabajando con complemento a dos y veinte bits, el uno es representado con el número (1x1024) = 1024 en notación hexadecimal 00400 y el menos uno con el número (-1x1024) = -1024 en notación hexadecimal FFC00.

De los diferentes algoritmos existentes para turbo decodificación en este trabajo se escogió el algoritmo MAX-Log-MAP con el fin de evitar las funciones no lineales y el gran número de operaciones de adición y multiplicación necesarias en el algoritmo MAP. El precio de esta simplificación es un detrimento de 0.5 dB en el desempeño [12].

**Turbo codificador**

El turbo codificador del estándar cdma2000 está conformado por dos codificadores constitutivos y un intercalador. El tamaño de paquete, *Nturbo*, usado en el turbo codificador fue de 250 bits. El número de símbolos codificados está dado por (*Nturbo*)/$R$+ 6/$R'$, donde $R$ es la tasa de podamiento, en este caso es de 1/2, con los patrones de podamiento establecidos por el estándar [9]. El turbo codificador también genera 6/$R'$ símbolos de cola de salida, estos bits de cola se obtienen después de que los codificadores constitutivos alcanzan el *Nturbo*, para este caso $R'$ que representa la tasa de podamiento de los bits de cola es de 1/2, es decir se generan 12 bits de cola.

El diseño del turbo codificador con sus datos de entrada, datos de salida, direcciones de intercalación y señales de control se presenta en la Figura 1.

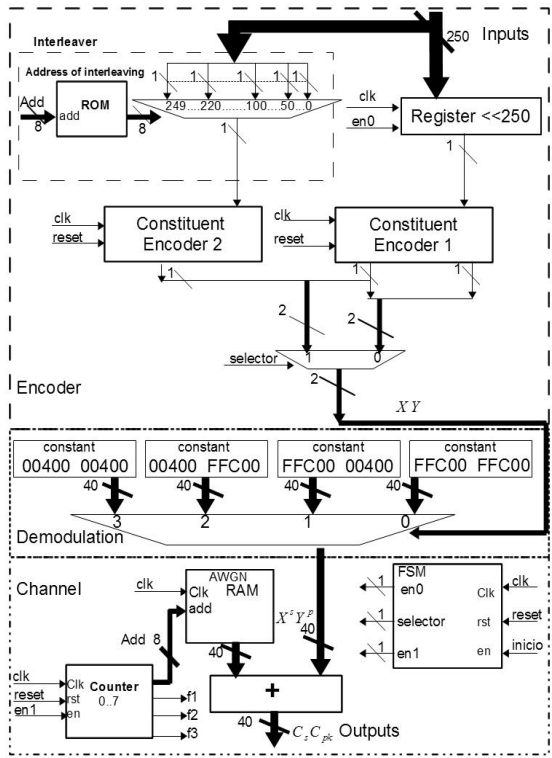

Figura 1. Turbo codificador.

En la Figura 1 se pueden apreciar tres bloques que están enmarcados por cuadros de línea punteada, el bloque superior corresponde al codificador, el bloque de la mitad al demodulador y el bloque inferior al canal.

**Bloque codificador**

Este bloque consta de dos unidades convolucionales y una unidad de turbo intercalación. Las unidades convolucionales están basadas en tres registros tipo D conectados de manera serial y cuatro compuertas lógicas XOR las cuales están conectadas de acuerdo a los polinomios generadores descritos en el estándar. La unidad de turbo intercalación está diseñada con una memoria ROM y un multiplexor, la memoria ROM contiene las direcciones de intercalación definidas por el algoritmo de intercalación. Para una trama de 250 bits, el multiplexor tiene 250 puertos de entrada de un bit, este módulo selecciona el bit de entrada que pasará a la segunda unidad convolucional, esta selección se hace por medio de la dirección de intercalación dada por la ROM.

En el bloque codificador, por cada bit de entrada el codificador genera una pareja de bits $XY$ en cada ciclo de máquina, donde $X$ es el bit sistemático y $Y$ el bit de paridad. El total de parejas $XY$ que genera el codificador es 256 (250 parejas de datos de información y seis parejas de datos de cola). Esta información ($XY$) pasa al bloque de demodulación.

**Bloque demodulador**

En el diseño hardware se partió de asumir los datos que entregaría un demodulador BPSK normalizados entre -1 y +1. Para esto se trabajó con un multiplexor de cuatro puertos, cuyas entradas corresponden al dato de uno y menos uno representados con las constantes 00400 y FFC00, las señales de selección vienen del codificador y están dadas por el bit sistemático ($X$) y el bit de paridad ($Y$). Los datos que entrega el demodulador son una pareja de 40 bits $X^s Y^p$ correspondientes al dato sistemático y de paridad.

**Bloque canal**

La unidad de canal está representada por una memoria RAM que contienen ruido blanco gausiano aditivo (*AWGN*), el cual es generado en matlab por la función "wgn", y un módulo de suma que permite contaminar con ruido los datos que vienen del demodulador ($X^s Y^p$), generando la salida $C_s C_{pk}$, donde $C_s$ es el dato de canal contaminado sistemático y $C_{pk}$ es el dato de canal contaminado de paridad, el subíndice $k$ es cero ó uno, indicando si es paridad cero o uno.

Los datos manejados en esta unidad son de tipo suave, entonces se normalizó los datos por 1024 para considerar tres cifras decimales; por ejemplo si se tiene un dato modulado de 1 y se le adiciona ruido de 2,256 la salida de canal es igual a 3,256 al normalizarlo por 1024 y truncarlo por piso queda 3334 en notación hexadecimal 00D06.

Dadas las técnicas de podamiento, las paridades que fueron podadas se reemplazan por cero, generando un total de datos de 759, los cuales son recibidos por el turbo decodificador, distribuidos en 253 datos sistemáticos, 506 datos de paridad discriminados en 253 de paridad uno y 253 de paridad cero. Los 253 datos sistemáticos son también intercalados generando otra entrada al turbo decodificador, en total recibe 253 parejas de datos de manera serial distribuidas alternadamente entre $C_S C_{p0}$, $C'_s C_{p1}$.

**Turbo decodificador**

Una gran dificultad de la implementación de los turbo códigos en muchas aplicaciones es la complejidad de la decodificación. El algoritmo MAP, el cual es un algoritmo óptimo de decodificación, no es tan apropiado a la hora de pensarse en una implementación en hardware, dado que requiere un gran número de operaciones de adición y multiplicación. Para disminuir esta complejidad, el algoritmo MAP puede ser transformado en el dominio logarítmico y a este nuevo algoritmo se le denomina Log-MAP [13].

Sin embargo al trabajar en implementaciones hardware en el dominio logarítmico, se hace necesario trabajar con el logaritmo Jacobiano, el cual expresa las funciones en términos de máximos más un factor de corrección, este factor puede ser almacenado en una tabla lo cual se traduce en hardware a bloques de memoria; una aproximación del algoritmo Log-MAP es no considerar este factor de corrección, con lo cual el cálculo de las funciones se reduce hallar la función máximo, a esta aproximación se le denomina algoritmo Max-Log-MAP.

Dado que en Max-Log-MAP no se tiene en cuenta los términos de corrección en las ecuaciones, las métricas hacia adelante ($\alpha$), hacia atrás ($\beta$) y la rata *LLR* para *m* estados, quedan expresadas de la siguiente forma:

$$\overline{\alpha}_k^{(m)} = \max_{m'} \{\alpha_{k-1}^{(m')} + \gamma_k(m', m)\} \quad (1)$$

$$\overline{\beta}_k^{(m)} = \max_{m'} \{\beta_{k+1}^{(m')} + \gamma_{k+1}(m', m)\} \quad (2)$$

$$L(d_k) = \max_m \{\alpha_k^{(m)} + \gamma_k^{(1,m)} + \beta_{k+1}^{f(1,m)}\} \\ - \max_m \{\alpha_k^{(m)} + \gamma_k^{(0,m)} + \beta_{k+1}^{f(0,m)}\} \quad (3)$$

Como se puede observar en las ecuaciones (1), (2) el cálculo de sus respectivas métricas en el tiempo *k* depende del valor de las métricas en los estados anteriores (*m´*) y de las ramas que conectan los estados anteriores con los estados actuales (*m*).

Otra observación importante es que en las ecuaciones anteriores, las operaciones que se deben realizar son sumas, comparaciones y restas, lo cual es menos complejo de implementar en hardware, entonces el algoritmo Max-Log-MAP establece un compromiso entre desempeño y complejidad ya que disminuye en 0.5 db su desempeño pero minimiza su complejidad de implementación.

Las condiciones de frontera para $\alpha$ en un tiempo $k = 0$ se dan en la ecuación (4) y para $\beta$ en un tiempo $k = \tau$, se dan en la ecuación (5).

$$\overline{\alpha}_0(0) = 0, \quad y \quad \overline{\alpha}_0(m) = -\infty \quad \forall \, m \neq 0 \quad (4)$$

$$\overline{\beta}_\tau(0) = 0, \quad y \quad \overline{\beta}_\tau(m) = -\infty \quad \forall \, m \neq 0 \quad (5)$$

En la Figura 2 se presenta el diagrama en bloques del turbo decodificador, el cual está basado en [14].

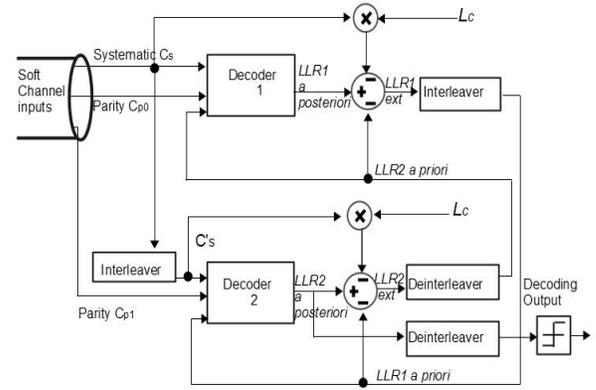

Figura 2. Turbo decodificador para cdma2000.

En la Figura 2 se puede observar el turbo decodificador propuesto para el codificador del estándar cdma2000, el cual tiene como entrada, los datos provenientes del codificador pasados por un canal *AWGN*, donde $C_s$ corresponde a los datos contaminados del canal para los bits sistemáticos (*X*), $C_{p0}$ y $C_{p1}$ a los datos contaminados del canal para los bits de paridad ($Y_0$ y $Y'_0$).

Según el estándar para *R*=1/2, sólo se tienen en cuenta las paridades $Y_0$ y $Y'_0$ se ignoran las paridades $Y_1$ y $Y'_1$.

Los turbo decodificadores se componen de dos decodificadores que usan el algoritmo MAX-Log-MAP, dos intercaladores y un deintercalador. Uno de los intercaladores es usado para intercalar la entrada sistemática $C_s$, la cual corresponde a una de las entradas del segundo decodificador y el otro intercalador se utiliza para intercalar la información extrínseca (*LLR ext*). Esta información es generada a partir de la tasa a posteriori (*LLR a posteriori*), la tasa a priori (*LLR a priori*) y los datos sistemáticos recibidos por el turbo decodificador $C_s$, esto se muestra en la ecuación (6).

$$LLR_{ext} = LLR_{a\,posteriori} - (LLR_{a\,priori} + C_s \times L_c) \quad (6)$$

El término *Lc* es la medida de confiabilidad del canal y es inversamente proporcional a la varianza y se define en (7), esta medida de confiabilidad tiene inferencia en el cálculo de la métrica de rama como se muestra en (8).

$$\sigma^2 = \frac{2a}{L_c} \quad (7)$$

$$\overline{\gamma}_k^{i,m} = \ln\left\{A_k \pi_k^i \exp\left[\frac{1}{\sigma^2}(C_{sk} u_k^i + C_{pk} v_k^{i,m})\right]\right\} \quad (8)$$

donde *a* es la ganancia del canal *AWGN*, la cual se define con valor de uno según [14], $\pi_k^i$ es la probabilidad a priori del bit de entrada, ($u_k, v_k$) son los bits de sistemático y de paridad transmitidos, ($C_s$, $C_{pk}$) son los datos recibidos en el decodificador, correspondientes a dato sistemático y de paridad.

Despejando la varianza de la ecuación (7) y reemplazándola en la ecuación (8), la métrica de rama queda expresada como:

$$\bar{\gamma}_k^{i,m} = \ln\left\{A_k \pi_k^i \exp\left[\frac{L_c}{2}(C_{sk} u_k^i + C_{pk} v_k^{i,m})\right]\right\} \quad (9)$$

$\pi_k^i$ se puede expresar en relación a la tasa *LLR a priori* ($L^e(u_k)$) en la ecuación (10).

$$\pi_k^i = B_k \exp\left[\frac{u_k^i L^e(u_k^i)}{2}\right] \quad (10)$$

Reemplazando la ecuación (10) en (9), la métrica de rama queda:

$$\gamma_k^{i,m} = \ln\left\{A_k B_k \exp\left[\frac{u_k^i L^e(u_k^i)}{2}\right]\exp\left[\frac{L_c}{2}(C_{sk} u_k^i + C_{pk} v_k^{i,m})\right]\right\} \quad (11)$$

Dado que los factores $A_k$ y $B_k$ son independientes de $u_k$, estos desaparecen al aplicar el logaritmo [15], luego, la ecuación (11) se expresa como:

$$\gamma_k^{i,m} = \frac{1}{2} u_k^i L^e u_k^i + \frac{1}{2} L_c C_{sk} u_k^i + \frac{1}{2} L_c C_{pk} v_k^{i,m} \quad (12)$$

Donde $L^e(u_k)$ es la *LLR a priori*, la cual proviene del deintercalador para $LLR_2$ *a priori* y del intercalador $LLR_1$ *a priori*, esto se puede observar en la Figura 2.

Para calcular las dos ramas en cada estado, se reemplaza $i$ por cero o uno en la ecuación (12) para todos los estados desde $m=0$ hasta $m=7$, teniendo en cuenta los bits transmitidos ($u_k, v_k$) se obtiene lo siguiente:

Tabla 1. Métricas de rama para el estándar cdma2000.

$$\gamma_k^{0,0} = \gamma_k^{0,1} = \gamma_k^{0,6} = \gamma_k^{0,7} = -\frac{1}{2}[LLR_{a\ priori} + L_c \times (C_s + C_p)]$$

$$\gamma_k^{1,0} = \gamma_k^{1,1} = \gamma_k^{1,6} = \gamma_k^{1,7} = \frac{1}{2}[LLR_{a\ priori} + L_c \times (C_s + C_p)]$$

$$\gamma_k^{0,2} = \gamma_k^{0,3} = \gamma_k^{0,4} = \gamma_k^{0,5} = -\frac{1}{2}[LLR_{a\ priori} + L_c \times (C_s - C_p)]$$

$$\gamma_k^{1,2} = \gamma_k^{1,3} = \gamma_k^{1,4} = \gamma_k^{1,5} = \frac{1}{2}[LLR_{a\ priori} + L_c \times (C_s - C_p)]$$

Como se observa en la Tabla 1, sólo es necesario calcular dos ramas para cada $k$, dado que las otras dos ramas son las opuestas, es decir, tienen signo contrario; de acuerdo a esto en la Figura 3 se presenta el trellis para el decodificador.

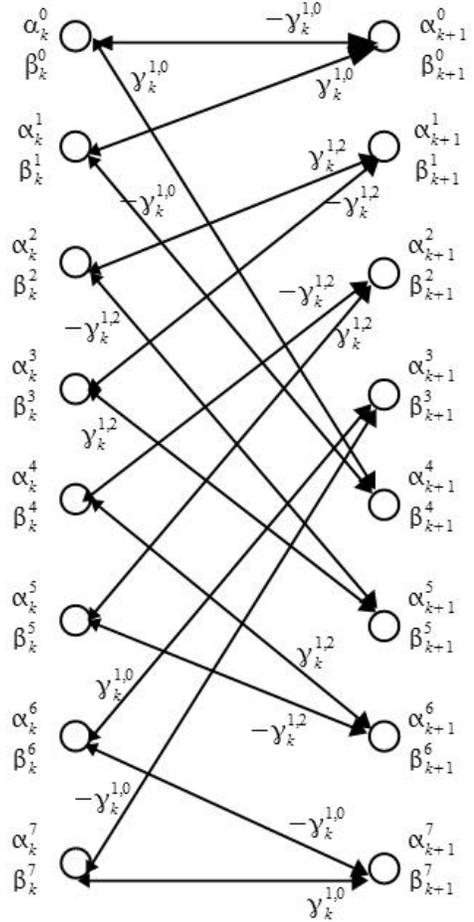

Figura 3. Trellis para el decodificador en etapa k.

En la Figura 3 se muestra el diagrama de trellis para cualquier tiempo $k$, se observa sólo dos ramas $\gamma_k^{1,0}$ y $\gamma_k^{1,2}$ como se mencionó en el párrafo anterior, distribuidas por todos los estados, para cada estado sale la rama y su opuesta aditiva.

Con el cálculo de las dos ramas por cada $k$, se simplifica el cálculo de las métricas hacia adelante ($\alpha$), hacia atrás ($\beta$) y la rata *LLR*, de acuerdo a las ecuaciones (1), (2) y (3) quedando expresadas de la siguiente manera:

$$\bar{\alpha}_{k+1}^{(m)} = \max_{m'}\left\{\left(\alpha_k^{(m')} + \gamma_k(m',m)\right), \left(\alpha_k^{(m'')} - \gamma_k(m',m)\right)\right\} \quad (13)$$

$$\bar{\beta}_k^{(m)} = \max_{m'}\left\{\left(\beta_{k+1}^{(m')} + \gamma_k(m',m)\right), \left(\beta_{k+1}^{(m'')} - \gamma_k(m',m)\right)\right\} \quad (14)$$

Las ecuaciones (13) y (14), las cuales se utilizan para el cálculo de las métricas hacia delante y hacia atrás para cada estado, están determinadas por la función máximo entre la suma y la resta. Para las métricas hacia adelante se trabaja con las métricas de la etapa anterior $k$ y para las métricas hacia atrás con las métricas de la etapa posterior $k+1$.

En la ecuación (15) se presenta la ecuación para hallar la rata LLR, esta ecuación se define como la resta entre la función máximo de las ramas de los unos y la de los ceros.

$$L(d_k) = \max \begin{Bmatrix} (\alpha_k^0 + \beta_{k+1}^4 + \gamma_k^{1,0}), (\alpha_k^1 + \beta_{k+1}^0 + \gamma_k^{1,0}), \\ (\alpha_k^2 + \beta_{k+1}^1 + \gamma_k^{1,2}), (\alpha_k^3 + \beta_{k+1}^5 + \gamma_k^{1,2}), \\ (\alpha_k^4 + \beta_{k+1}^6 + \gamma_k^{1,2}), (\alpha_k^5 + \beta_{k+1}^2 + \gamma_k^{1,0}), \\ (\alpha_k^6 + \beta_{k+1}^3 + \gamma_k^{1,0}), (\alpha_k^7 + \beta_{k+1}^7 + \gamma_k^{1,0}) \end{Bmatrix}$$
$$- \max \begin{Bmatrix} (\alpha_k^0 + \beta_{k+1}^0 - \gamma_k^{1,0}), (\alpha_k^1 + \beta_{k+1}^4 - \gamma_k^{1,0}), \\ (\alpha_k^2 + \beta_{k+1}^5 - \gamma_k^{1,2}), (\alpha_k^3 + \beta_{k+1}^1 - \gamma_k^{1,2}), \\ (\alpha_k^4 + \beta_{k+1}^2 - \gamma_k^{1,2}), (\alpha_k^5 + \beta_{k+1}^6 - \gamma_k^{1,2}), \\ (\alpha_k^6 + \beta_{k+1}^7 - \gamma_k^{1,0}), (\alpha_k^7 + \beta_{k+1}^3 - \gamma_k^{1,0}) \end{Bmatrix} \quad (15)$$

Para el diseño del turbo decodificador se tuvo en cuenta el de la Figura 2, con la diferencia que se diseñó un sólo decodificador, dado que el segundo decodificador inicia el proceso cuando el primer decodificador haya terminado de procesar todos sus datos. En la figura 4 se presenta el diagrama en bloques del turbo decodificador.

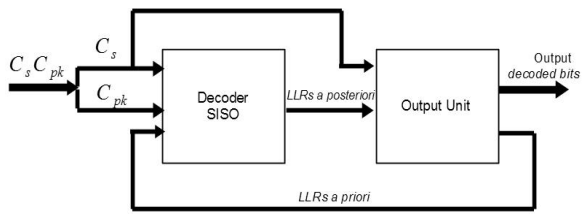

Figura 4. Diagrama en bloques del turbodecodificador.

En la Figura 4 se puede observar que el turbo decodificador consta de dos unidades básicas: unidad de decodificador SISO y unidad de salida.

En la unidad SISO se realiza el cómputo del decoder1 o del decoder 2 como se presentó en la figura 2 y en la unidad de salida se realizan una serie de tareas para generar la señal de realimentación (LLR a priori) para el decoder SISO y la salida de los bits decodificados.

En la Figura 5 se presenta el algoritmo de turbo decodificación, describiendo todas las tareas que se deben realizar. Se debe tener en cuenta que sólo se utiliza un decodificador SISO como se planteo en la Figura 4.

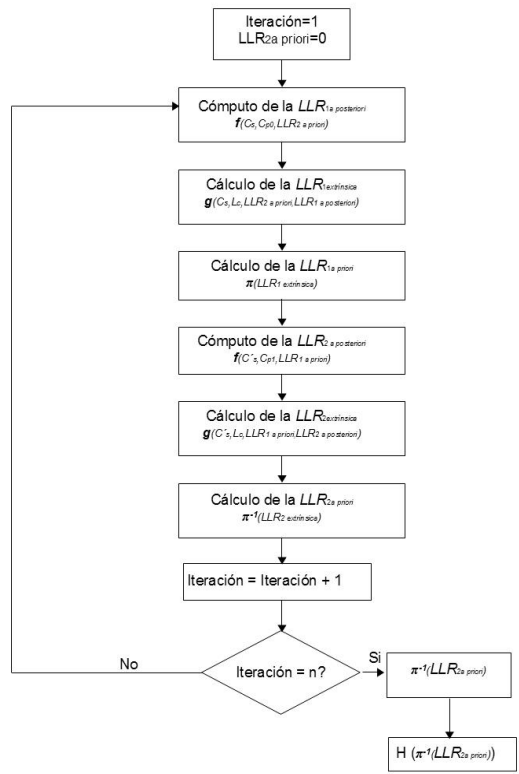

Figura 5. Diagrama de flujo del proceso de turbo decodificación.

En la Figura 5 se muestra el proceso de turbo decodificación, primero se inicia el cómputo del decodificador (decoder 1) y después el decodificador (decoder 2) y se van alternando hasta terminar todas las iteraciones, una vez terminado este proceso se genera la salida con los bits decodificados.

Hay que tener presente que la información que se maneja en cada iteración es suave, mientras que la salida decodificada es dura, es decir uno o cero por cada bit decodificado.

### Cómputo *Decoder* 1

Inicialmente no hay realimentación por lo tanto la tasa $LLR_{2\ a\ priori}$ es cero, y se da inicio al cómputo de la $LLR_{1\ a\ posteriori}$, este cálculo es la función $f(C_s, C_{p0}, LLR_{2\ a\ priori})$ cuyos parámetros son los datos de canal (sistemático y paridad cero) y la rata de realimentación, esta función $f$ es la rata $LRR$ ($L(d_k)$) que corresponde a la ecuación (15) y es la salida del decoder SISO como se observó en la Figura 4.

Después se halla la $LLR_{1\ extrínseca}$ por medio de la función $g(C_s, L_c, LLR_{2a\ priori}, LLR_{1a\ posteriori})$, la función $g$ corresponde a la ecuación (6).

El siguiente paso es hallar la rata $LLR_{1\ a\ priori}$ aplicando la función de intercalación $\pi$ a la rata $LLR_{1\ extrínseca}$, este es el último paso del cómputo del decodificador (decoder 1).

**Cómputo *Decoder* 2**

Este inicia calculando la $LLR_{2\ a\ posteriori}$ por medio de la función $f(C'_s, C_{p1}, LLR_{1\ a\ priori})$, con la diferencia que la realimentación para este caso ya no es cero sino la rata $LLR_{1\ a\ priori}$ obtenida en el cómputo del decodificador (decoder 1) y los datos $C'_s$ son los datos sistemáticos que recibe el turbo decodificador pero que han sido intercalados y $C_{p1}$ los datos de paridad uno.

Después se halla la rata $LLR_{2\ extrínseca}$ a través de la función $g(C'_s, L_c, LLR_{1a\ priori}, LLR_{2\ a\ posteriori})$ y finalmente se halla la rata $LLR_{2\ a\ priori}$ aplicando la función de deintercalación $\pi^{-1}$ a la rata $LLR_{2\ extrínseca}$ y se termina el cómputo del *decoder* 2.

Una vez se ha realizado el cómputo de los decodificadores (*decoder* 1 y 2) se termina la iteración, y continua el proceso hasta que se hagan las *n* iteraciones.

Después de que se haya concluido las *n* iteraciones se aplica la función de deintercalación $\pi^{-1}$ a la rata $LLR_{2\ a\ priori}$ y se realiza la función de decisión dura *H* para obtener los bits de salida decodificados, la función *H* se especifica a continuación.

$$H = \begin{cases} Si\ LLR_{2\ a\ priori}\ \pi^{-1} < 0 \Rightarrow el\ bit\ decodificado = 0 \\ Sino\ el\ bit\ decodificado = 1 \end{cases}$$

En la Figura 6 se presenta el diseño en hardware del turbo decodificador, para una entrada serial de 253 parejas de datos (sistemático, paridad) y una salida de 250 bits decodificados.

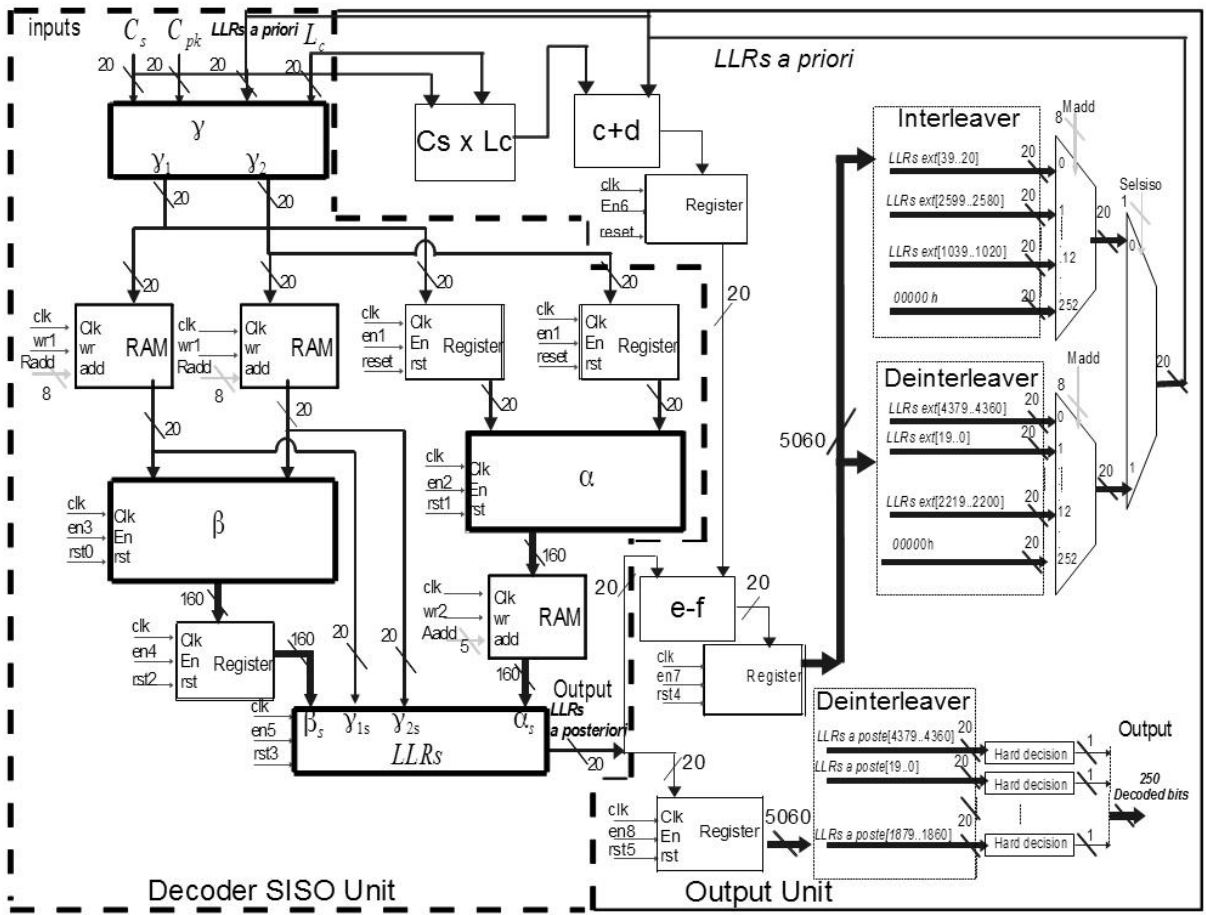

Figura 6. Arquitectura de la unidad de turbodecodificación.

En la Figura 6 se pueden ver dos grandes unidades, la del decodificador SISO que se encuentra enmarcada en el cuadro de línea punteada y la unidad de salida que se encuentra dentro del cuadro de línea continua.

Todas las señales de control que tienen las dos unidades del turbo decodificador son manejadas por una unidad de control implementada sobre una máquina de estados finitos.

**Unidad de decodificador SISO**

En la Figura 6, se observa que el decodificador SISO está constituido por cuatro bloques: cálculo de ramas ($\gamma$), cálculo de métricas hacia adelante ($\alpha$), cálculo de métricas hacia atrás ($\beta$) y el bloque del cálculo de *LLRs*.

**Bloque de cálculo de ramas**

En este bloque se hace el cálculo de ramas; de acuerdo a la Tabla 1, se implementó sólo el cálculo de dos ramas por cada dato $\gamma_k^{1,0}$ y $\gamma_k^{1,2}$. Estas ramas se almacenan en registros para ser usadas inmediatamente en el cómputo de las métricas hacia adelante, y también se van almacenando en memorias para ser utilizadas después en el bloque ($\beta$) y en el de *LLRs*. En la Figura 7 se muestra la unidad básica del bloque.

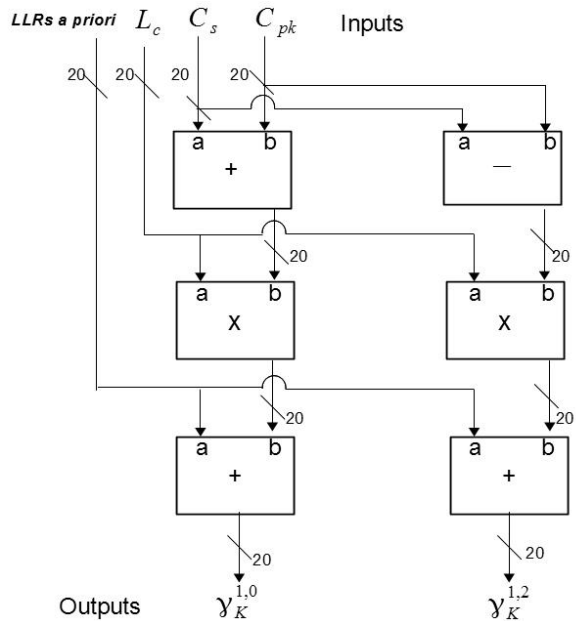

Figura 7. Bloque unidad cálculo de rama.

Como se muestra en la Figura 7, para el cálculo de ramas sólo es necesario bloques muy básicos que realicen las operaciones de suma resta y multiplicación entre los datos de entrada LLRs a priori, LC y los datos sistemáticos y de paridad recibidos en el decodificador.

**Bloque cálculo de métricas hacia adelante (α)**

Este bloque es el encargado de realizarse el cálculo de las métricas hacia delante para un estado $m$ en un tiempo $k+1$ y depende de los valores de dos estados $m'$ y $m''$ en el tiempo anterior $k$.

Por cada dato se tienen ocho estados, entonces se calcula la métrica para cada estado obteniendo una salida de ciento sesenta bits, esta información se almacena en un registro para ser usada en el cálculo de la siguiente métrica del próximo dato, de igual manera estas métricas se almacenan en una memoria, para posteriormente ser usadas en el bloque de *LLRs*. En la Figura 8 se presenta este bloque.

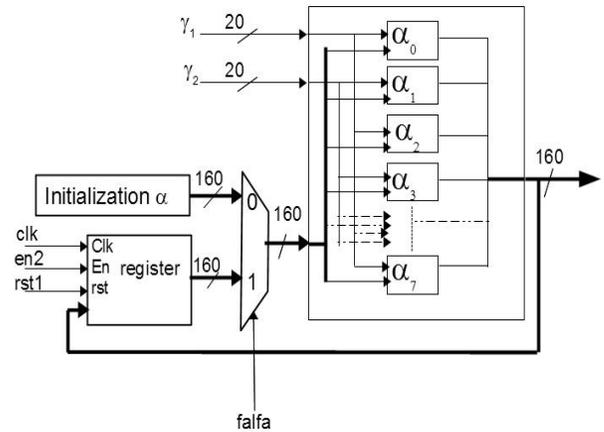

Figura 8. Bloque de los alfa.

En la Figura 8 se observa que este bloque esta conformado por ocho bloques alfa (α), que son los que realizan el cómputo de la métrica para cada estado, un registro que guarda los alfas calculados y un bloque de inicialización para los alfas como se expresó en la ecuación (4).

Los valores de inicialización para los estados de uno a siete se utilizan valores muy negativos, estos se expresaron con un valor de -250 y para el estado cero un valor de cero. Estos valores se presentan en base hexadecimal y teniendo en cuenta la normalización de 1024 y el complemento a dos, queda expresada la inicialización de los ocho estados de la siguiente manera:

C1800,C1800,C1800,C1800,C1800,C1800,C1800,00000.

En la Figura 9 se presenta la celda básica de los bloques alfa.

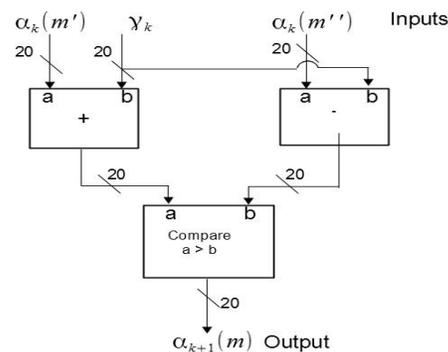

Figura 9. Celda cálculo de métrica hacia delante.

Como se observa en la Figura 9 el bloque alfa está constituido por bloques básicos, sumador, restador y un comparador de mayor, como se describe en la ecuación (13).

**Bloque cálculo de métricas hacia atrás (β)**

Este bloque inicia su cálculo de métrica hacia atrás una vez se haya calculado las 253 ramas $\gamma_1$ y $\gamma_2$. Al igual que el bloque α, tiene 8 bloques de cálculo de métricas beta (β), un bloque de inicialización de acuerdo a la ecuación (5), y una salida de ciento sesenta bits la cual es almacenada en un registro, para ser usada en el cálculo de la siguiente métrica del próximo dato. En la Figura 10 se presenta la unidad básica del bloque que calcula la métrica hacia atrás β.

En la Figura 10 se muestra el bloque de la métrica hacia atrás, el cual tiene la misma celda básica que α, pero con diferentes entradas y está basada en la ecuación (14).

Esta métrica tiene en cuenta para su cálculo el valor de los estados ($m'$ y $m''$) en el tiempo $k+1$, al igual que las ramas que conectan los estados de $k+1$ con el estado ($m$) que se va hallar en el tiempo $k$.

**Bloque cálculo de *LLRs***

Este bloque calcula las ratas *LLRs* de acuerdo a la ecuación (15) y guarda el resultado en un registro. Las entradas de este bloque provienen de las memorias que contienen las ramas, la memoria de las métricas (α) y el registro de las métricas (β), este bloque inicia calculando el último *LLR*, dado que las métrica hacia atrás entregan inicialmente la última métrica. En la Figura 11 se presenta este bloque.

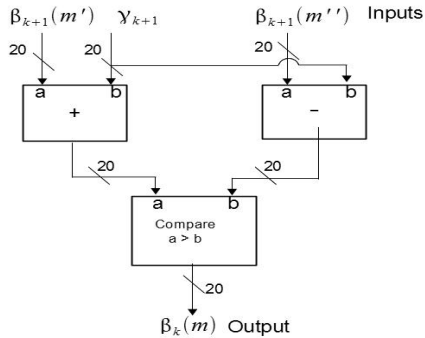

Figura 10. Celda bloque de métrica hacia atrás.

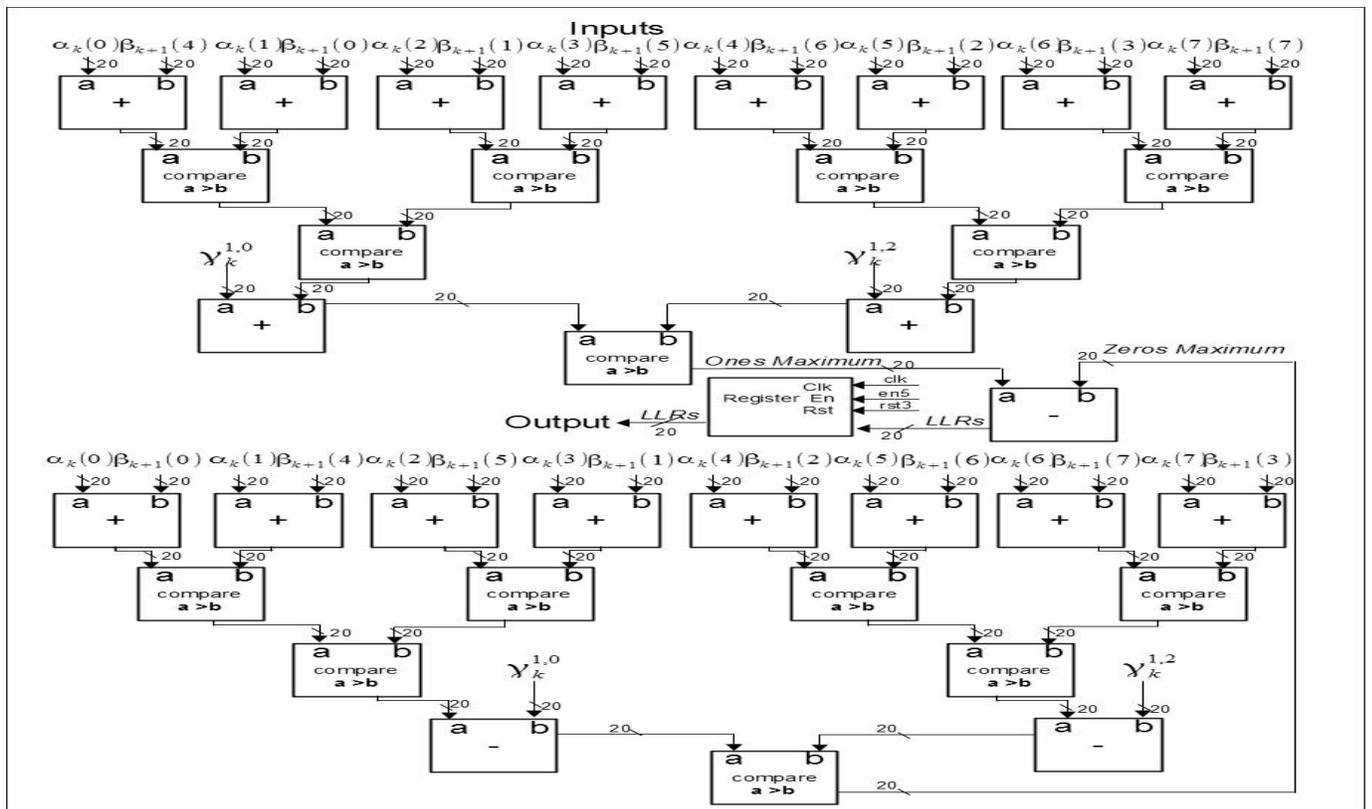

Figura 11. Bloque cálculo de la rata *LLRs*.

En la Figura 11 se observa que el bloque de *LLRs* está constituido por sumadores, restadores y comparadores de mayor, lo cual constituye las funciones de máximo para los unos y ceros como se describe en la ecuación (15).

**Unidad de salida**

En la Figura 6 se muestra está unidad la cual tiene como tarea entregar los datos decodificados. Está unidad consta de bloques de intercalación y deintercalación, registros y bloques aritméticos como sumador, restador y multiplicador.

Los bloques de intercalación y deintercalación, reciben los datos de un registro de desplazamiento de 5060 bits que corresponden a 253 datos, estos se distribuyen de manera paralela teniendo en cuenta la dirección de intercalación o deintercalación, sobre dos multiplexores de 253 puertos, de los cuales se selecciona el dato intercalado y deintercalado, pasando a un tercer multiplexor de dos puertos, en donde la señal de selección si está en cero, deja pasar los datos intercalados o en uno los datos deintercalados, de acuerdo al algoritmo planteado en la Figura 5.

En los bloques aritméticos se hace el cálculo de la *LLR extrínseca* (*LLR* ext), basado en la ecuación (6), este resultado se va almacenando en un registro de desplazamiento hasta completar las 253 *LLR extrínsecas*.

Finalmente estas *LLR extrínsecas* pasan a hacer intercaladas o deintercaladas, generando los datos de realimentación del SISO (*LLR a priori*) para la siguiente iteración.

La salida de los 250 bits decodificados son tomados de los datos entregados por el SISO (*LLR $_{a\ posteriori}$*) y son almacenados en un registro de desplazamiento, los cuales son posteriormente deintercalados y comparados con cero (decisión dura) de acuerdo a la función *H*, para generar la salida final, esto se hace de manera simultánea ya que se tienen 250 comparadores.

**Unidad de control**

En la Figura 12 se presenta la máquina de estados implementada para la unidad de control.

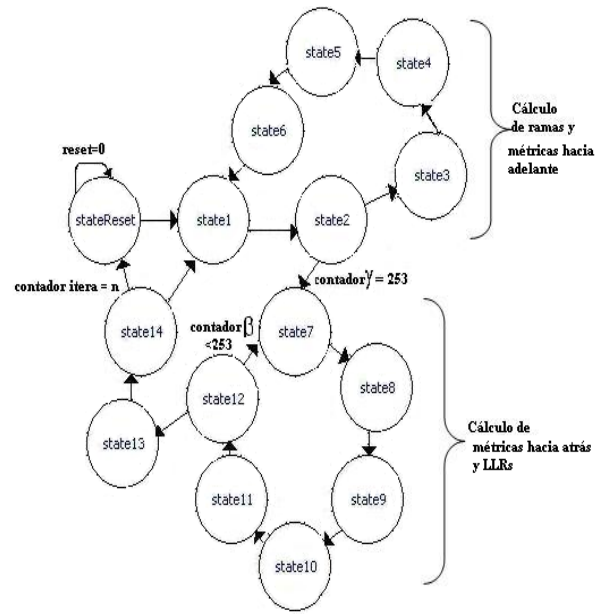

Figura 12. Máquina de estados de la unidad de control

Como se observa en la Figura 12 fueron necesarios 14 estados y el estado de inicialización o reset. De estos catorce, los seis primeros (state1-state6) se utilizan para el cálculo de las ramas y métricas hacia adelante, los siguientes 6 estados (state7-state12) se utilizan para el cálculo de las métricas hacia atrás y las tasas *LLRs* y los últimos dos estados (state13,state14) para incrementar y evaluar el contador de iteraciones, cuando este contador llega a la *n* iteraciones pasa al estado de inicio (stateReset).

El cálculo de la turbo decodificación se debe hacer para 253 datos de canal, por lo tanto el número de veces que toma realizar el cómputo de ramas, métricas hacia adelante, métricas hacia atrás y *LLRs* es de 253, para esto es necesario tener un contador de ramas (contadorγ) y un contador de métricas hacia atrás (contadorβ), que cuando lleguen a 253 respectivamente salen del ciclo, esto se muestra en la Figura 12 donde se ve claramente los dos ciclos (state1-state6) y (state7-state12).

## RESULTADOS

En esta sesión se presenta los resultados obtenidos para el turbo decodificador y el análisis de desempeño para el diseño propuesto.

En las Tablas 2,3 y 4 se presentan los resultados obtenidos para el turbo decodificador teniendo en cuenta tres diferentes relaciones de señal a ruido (SNR) baja, media y alta, donde una SNR baja significa un alto grado de contaminación de los datos por el canal, para las pruebas se considero un valor de 0.35, una SNR media de 1.35 significa un grado medio de contamina-ción y una SNR baja de 2.35 hace referencia a un mínimo grado de contaminación del canal sobre los datos. Para las pruebas se consideraron 7 iteraciones, donde cada iteración involucra el procesamiento del SISO dos veces (para el procesamiento de los datos intercalados y los deintercalados).

Tabla 2 Resultados de decodificación para una SNR=0.35

|  | Bits decodificados correctamente |
|---|---|
| 1 iteración | 189 |
| 2 iteración | 189 |
| 3 iteración | 190 |
| 4 iteración | 180 |
| 5 iteración | 180 |
| 6 iteración | 182 |
| 7 iteración | 187 |

Tabla 3 Resultados de decodificación para una SNR=1.35

|  | Bits decodificados correctamente |
|---|---|
| 1 iteración | 217 |
| 2 iteración | 238 |
| 3 iteración | 247 |
| 4 iteración | 250 |
| 5 iteración | 250 |
| 6 iteración | 250 |
| 7 iteración | 250 |

Tabla 4 Resultados de decodificación para una SNR=2.35

|  | Bits decodificados correctamente |
|---|---|
| 1 iteración | 250 |
| 2 iteración | 250 |
| 3 iteración | 250 |
| 4 iteración | 250 |
| 5 iteración | 250 |
| 6 iteración | 250 |
| 7 iteración | 250 |

Como se observa en la Tabla 2 no se logra la decodificación total del paquete, esto se debe a que la contaminación de ruido es tan alta que el decodificador no logra su objetivo, además el proceso de la realimentación tiene un efecto negativo debido a que en cada iteración se contaminan más los datos, dado que si el número de errores es alto con la realimentación se introduce más error, lo que hace que no mejore la decodificación.

En la Tabla 3 se ve como el efecto de realimentación en cada iteración mejora el proceso de decodificación, además la SNR está en el rango medio. Y en la Tabla 4 se observa que el SISO logra decodificar los 250 bits desde la primera iteración.

De acuerdo a las Tablas 3 y 4 podemos ver que es suficiente con 4 iteraciones para la decodificación.

En la Figura 13 se muestra las curvas de BER vs $E_b/N_0$ para los resultados del turbo decodificador con una entrada de 250000 bits (1000 paquetes de 250 bits).

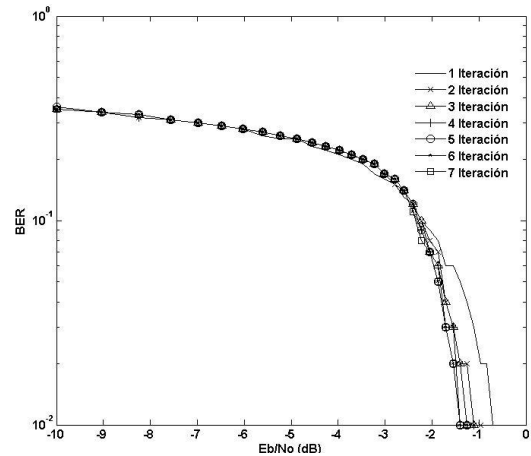

Figura 13. Curvas de BER vs $E_b/N_0$.

En la Figura 13 se puede observar que a partir de -2.5dB la curva de la iteración 7 presenta una menor relación de Eb/No; también se puede apreciar que la rata de error de bit (BER) disminuye por cada iteración. La BER más baja alcanzada está en el orden de $10^{-2}$, esto se debe al tamaño de la muestra de entrada que está en 250000 bits, si se desea obtener una BER del orden de $10^{-7}$ se necesita una entrada mínima de 10 millones de bits es decir cuarenta mil paquetes de 250 bits.

Los resultados obtenidos en área y velocidad al sintetizarlo en la FPGA Stratix III EP3SE80F1152C2 se presentan en la Tabla5.

Tabla 5 Parámetros de área y desempeño

|  | AREA (ALUTs) | VELOCIDAD (MHz) |
|---|---|---|
| Turbo decodificador | 6367 | 101.97 |

De acuerdo a los resultados presentados en la Tabla 5 se alcanza una frecuencia de reloj de 101.97MHz.

Se debe tener en cuenta que el número de ciclos para cada iteración se debe multiplicar por dos, dado que el sistema para cada iteración hace intercalación y después deintercalación.

El número de ciclos que gasta en una sola iteración para un paquete de 253 bits (incluidos bits de cola) es de 6081 ciclos esto se puede obtener de la siguiente ecuación:

$$Ciclos = Inicialización + \{Cálculo\ ramas\ y\ alfas + Cálculo\ Betas\ y\ LLRs\} * Número\ de\ iteraciones * 2 \quad (16)$$

```
Ciclos  = 1 + {{[ 6 * packetSize ] + 2} +
{[ 6 * packetSize ] + 2}} * NIter * 2

Ciclos  = 1 + {{[ 6 * 253 ] + 2} +
{[ 6 * 253 ] + 2}} * 1 * 2 = 6081
```

El tiempo total para la primera iteración es el número de ciclos por periodo (6081* 9,807ns) 59,636 $\mu$s.

El tiempo total para 4 iteraciones en donde ya se tienen resultados de la decodificación para SNR bastante altas, es de 238,516 $\mu$s, luego la tasa que está dada por la relación tamaño de paquete sobre el tiempo, es de 1,048Mbits/s. En el caso de SNR bajas sólo basta con una sola iteración como se observó en la tabla 4, para este caso la tasa obtenida es de 4,19Mbits/s.

Las tasas alcanzadas no son muy altas debido a que el diseño es secuencial y va procesando dato a dato como se observó en la ecuación (16).

Dado el retardo que tiene el sistema secuencial se desarrolló otro diseño en el cual se paraleliza el sistema, este diseño consiste en realizar de manera simultanea el cálculo para 11 datos, para este caso el paquete de 253 incluida la información de colas se divide en 11 datos de procesamiento paralelo para un total de 23 paquetes, entonces en la ecuación (16) el tamaño de paquete es de 23.

Con esta paralelización se obtiene que para una iteración el número de ciclos es 561. Sin embargo la frecuencia obtenida descendió considerablemente a 25 MHz, y el área ocupada fue de 28085 ALUTs, es decir, se cuadruplico en área y se redujo en velocidad la cuarta parte. Esto se debe a que las unidades para el cálculo de ramas, métricas hacia delante, métricas hacia atrás y *LLRs*, ahora realizan cálculos de manera simultánea para 11 datos, es decir que los bloques que se presentaron en las Figuras del 7 al 11 se replican once veces, lo que aumenta considerablemente el área.

Pero esto no afecta finalmente el desempeño, dado que con 561 ciclos para la primera iteración por el periodo de 40ns, se obtiene un tiempo total para la 1 iteración de 22,44$\mu$s. Entonces la tasa obtenida para 4 iteraciones es de 2,78Mbits/s y para 1 iteración es de 11,14Mbits/s, esta tasas corresponden al doble de las obtenidas en el caso serial, exactamente a 2.6 veces.

## CONCLUSIONES

De acuerdo a los resultados obtenidos para las tasas de una o cuatro iteraciones se ve claramente que el paralelismo juega un papel muy importante a la hora del cálculo del tiempo total de ejecución.

A pesar de que la frecuencia se redujo a la cuarta parte esto no afecta el resultado de la tasa, ya que el tamaño del paquete de procesamiento como se indica en la ecuación (16) es un factor multiplicativo que hace que el número de ciclos se reduzca en un factor de 10 para el caso paralelo.

La naturaleza serial del algoritmo Max-Log-MAP hace que sea difícil una implementación paralela en hardware; sin embargo, se realizó paralelismo a nivel de procesamiento de los datos, como fue el caso de generar, de manera simultánea el cálculo de ramas, alfas, betas y *LLRs* para 11 datos.

Una desventaja evidente de los turbo códigos es el retardo de decodificación causado por el número de iteraciones que se requieren para una BER dada. Sin embargo cuando la SNR es suficientemente buena, es decir que el ruido de contaminación no es tan alto, el proceso de decodificación converge al valor de mayor corrección posible, que el decodificador puede entregar en un número menor de iteraciones, lo cual coincide exactamente con lo que predice la teoría de turbo codificación.

Para la obtención de datos de BER prácticos se debe tener en cuenta que el número de bits que se necesitan para lecturas de BER muy bajas puede ser muy alto, ya que se puede requerir un número de paquetes de prueba grande para obtener un modelamiento correcto de BER lo cual trae como consecuencia un tiempo de simulación y procesamiento bastante alto. Por Ejemplo: una BER de $1\times10^{-7}$ significa que 1 de cada 10 millones de bits es erróneo. Por teoría básica de estadística para poder asegurar que una BER es de $1\times10^{-7}$ se necesitan por lo menos 10 pruebas donde se cumpla la condición. Esto significa que se necesitarían enviar por lo menos 100 millones de bits (para paquetes de 250 bits, esto es 400.000 paquetes), lo cual se prueba a través de un tiempo de procesamiento bastante largo.

## REFERENCIAS